\documentclass[10pt,conference]{IEEEtran}
\IEEEoverridecommandlockouts  
\usepackage{cite}
\usepackage{amsmath,amssymb,amsfonts}
\usepackage{algorithmic,balance,makecell}
\usepackage{graphicx}
\usepackage{textcomp}
\usepackage{xcolor}
\usepackage{booktabs}
\usepackage{algorithm}
\usepackage{multirow}
\usepackage{xspace}
\usepackage{enumitem}
\def\BibTeX{{\rm B\kern-.05em{\sc i\kern-.025em b}\kern-.08em
    T\kern-.1667em\lower.7ex\hbox{E}\kern-.125emX}}

\newcommand\toolname{SWE-Doctor\xspace}

\usepackage[framemethod=tikz]{mdframed}
\newcounter{finding}
\newcommand{\finding}[1]{\refstepcounter{finding}
  \vspace{2.3mm}
 \begin{mdframed}[linecolor=gray,roundcorner=12pt,backgroundcolor=gray!15,linewidth=3pt,innerleftmargin=2pt, leftmargin=0cm,rightmargin=0cm,topline=false,bottomline=false,rightline = false]
  \textbf{Finding \arabic{finding}:} #1
 \end{mdframed}
 \vspace{2.3mm}
}

\makeatletter
\long\def\@makecaption#1#2{%
\ifx\@captype\@IEEEtablestring%
\footnotesize\bgroup\par\centering\@IEEEtabletopskipstrut{\normalfont\footnotesize {#1.}\nobreakspace #2}\par\addvspace{0.5\baselineskip}\egroup%
\@IEEEtablecaptionsepspace
\else
\@IEEEfigurecaptionsepspace
\setbox\@tempboxa\hbox{\normalfont\footnotesize {#1.}\nobreakspace #2}%
\ifdim \wd\@tempboxa >\hsize%
\setbox\@tempboxa\hbox{\normalfont\footnotesize {#1.}\nobreakspace}%
\parbox[t]{\hsize}{\normalfont\footnotesize \noindent\unhbox\@tempboxa#2}%
\else%
\hbox to\hsize{\normalfont\footnotesize\hfil\box\@tempboxa\hfil}%
\fi\fi}
\makeatother

\begin{document}

\title{SWE-Doctor: Guiding Software Engineering Agents with Runtime Diagnosis from Multi-Faceted Bug Reproduction Tests}

\author{\IEEEauthorblockN{Yaoqi Guo\IEEEauthorrefmark{1},
Yang Liu\IEEEauthorrefmark{1},
Jie M. Zhang\IEEEauthorrefmark{2}, 
Yun Ma\IEEEauthorrefmark{3},
Yiling Lou\IEEEauthorrefmark{4}, 
Zhenpeng Chen\IEEEauthorrefmark{5}}
\IEEEauthorblockA{\IEEEauthorrefmark{1}Nanyang Technology University, \IEEEauthorrefmark{2}King's College London, \IEEEauthorrefmark{3}Peking University, \IEEEauthorrefmark{4}University of Illinois Urbana-Champaign,} \IEEEauthorblockA{\IEEEauthorrefmark{5}Tsinghua University}
\IEEEauthorblockA{yaoqi001@e.ntu.edu.sg, yangliu@ntu.edu.sg, jie.zhang@kcl.ac.uk, mayun@pku.edu.cn, yilingl@illinois.edu,} \IEEEauthorblockA{zpchen@tsinghua.edu.cn}
\thanks{Corresponding author: Zhenpeng Chen.}}

\pagestyle{plain} 
\maketitle
\thispagestyle{plain} 

\begin{abstract}
Large language model (LLM)-based software engineering agents are increasingly developed to resolve software issues by generating patches from issue reports and code repositories. Bug reproduction tests (BRTs) are an important building block for such agents and have been shown useful for patch validation. However, it remains unclear whether BRTs can also help the more central stage of patch generation. We first conduct a preliminary study and find that directly using advanced BRT generators to guide patch generation is not beneficial: fail-to-fail BRTs can mislead agents, while even fail-to-pass BRTs bring limited or negative gains. Our analysis reveals two reasons: fail-to-pass BRTs may cover only one manifestation of the reported issue, leading to partial patches, whereas fail-to-fail BRTs are unreliable as direct patch-generation targets. Motivated by these insights, we propose \toolname, a software issue resolution agent that guides patch generation with runtime diagnoses derived from multi-faceted BRT executions. \toolname first generates multi-faceted BRTs for different behavioral requirements stated in the issue, then executes and debugs these BRTs to construct runtime-grounded diagnosis records, and finally uses the diagnoses together with localization information inferred during BRT generation to guide patch generation and reduce partial patches. We evaluate \toolname on Python bug-fixing issues from the widely adopted SWE-bench Verified and SWE-bench Pro across five LLM backends. \toolname consistently outperforms existing agents across all 10 LLM--benchmark combinations, achieving average resolution rates of 75.7\% on SWE-bench Verified and 59.4\% on SWE-bench Pro. In particular, on the more challenging SWE-bench Pro, \toolname improves the average resolution rate by 8.0--8.9 percentage points over the baseline agents. Further analyses show that both multi-faceted BRT generation and runtime-grounded diagnosis contribute substantially to \toolname's effectiveness.
\end{abstract}


\section{Introduction}\label{sec:introduction}
Large language models (LLMs) have accelerated the development of software engineering agents for resolving real-world software issues~\cite{sweagent, agentless, swebench}. Given a reported issue and its corresponding buggy repository, such agents can navigate the codebase, inspect relevant files, edit source code, run tests, and submit patches in an end-to-end manner, with LLMs serving as their reasoning and decision-making backends.

Bug reproduction test (BRT) generation is the task of generating tests that reproduce the bugs described in issue reports~\cite{eotterpp}. It is an important building block for software issue resolution because a BRT turns a textual bug report into executable feedback~\cite{brtagent,sweagent}. In real-world issue resolution tasks, however, such tests are usually not provided~\cite{agentless}: an agent must infer the bug-triggering behavior from the issue report and the repository itself. As a result, effective BRT generation can provide agents with concrete executions that expose the reported bug and support downstream resolution.

Recent studies~\cite{eotterpp, issue2test, assertflip} have proposed advanced BRT generators and shown that generated BRTs can improve issue resolution when used for patch validation, i.e., checking whether a generated patch passes the BRT. However, patch validation happens only after a candidate patch has already been produced. It remains unclear whether BRTs can also help the more central and challenging stage of patch generation itself. To investigate this question, we conduct a preliminary study that integrates three state-of-the-art BRT generators into the patch-generation process of a widely used issue resolution agent. Surprisingly, the results show that advanced BRT generators do not improve overall patch generation.

Our analysis reveals two reasons. First, fail-to-pass BRTs, i.e., tests that fail on the original repository but pass on the patched repository, often cover only one manifestation of the reported issue, causing the agent to produce partial patches that satisfy the generated test while leaving other stated behaviors unaddressed. This motivates us to generate BRTs that cover multiple facets of the issue. Second, fail-to-fail BRTs, i.e., tests that fail on both the original and patched repositories, can mislead the agent when used as direct patch-generation targets. This suggests that patch generation should not treat BRTs merely as end-to-end pass/fail targets, but should make more effective use of the information exposed by BRT executions.

Inspired by these findings, we propose \toolname, a software issue resolution agent that guides patch generation by generating multi-faceted BRTs and converting their executions into runtime diagnoses. \toolname first extracts multi-faceted behavioral requirements from the issue report and generates targeted BRTs for them, so that different facets of the reported behavior can be exercised rather than relying on a single narrow test. It then executes and debugs these BRTs to construct runtime-grounded diagnosis records that summarize patch-relevant observations, such as suspected fault locations, failure symptoms, propagation paths, and runtime values. Finally, \toolname uses these diagnosis records, together with localization information inferred during BRT generation, to guide patch generation and adds a completeness check before submission to reduce partial patches.

We evaluate \toolname on Python bug-fixing issues from the widely used SWE-bench Verified benchmark~\cite{swebench} and the more challenging SWE-bench Pro benchmark~\cite{swebenchpro}, using five LLM backends from four vendors. We compare \toolname against two representative baselines: mini-SWE-agent~\cite{sweagent}, a widely used research baseline, and live-SWE-agent~\cite{liveswe}, a recently proposed agent with leading performance on SWE-bench-style tasks. \toolname achieves average resolution rates of 75.7\% on SWE-bench Verified and 59.4\% on SWE-bench Pro, and consistently outperforms both baselines across all 10 LLM--benchmark combinations. The gains are especially pronounced on the challenging SWE-bench Pro, where \toolname improves the average resolution rate by 8.0--8.9 percentage points over the baselines. Moreover, \toolname uniquely resolves more than twice as many issues as either baseline, suggesting that it expands the set of issues that current agents can resolve.
 
Ablation results show that both multi-faceted BRT generation and runtime-grounded bug diagnosis contribute substantially: removing either stage significantly reduces the issue resolution rate, while the ablated variants still outperform the baselines. A finer-grained analysis shows that the two stages contribute in different ways. Multi-faceted BRT generation mainly improves resolution on issues where the generated BRTs are fail-to-pass tests, whereas runtime-grounded diagnosis mainly improves resolution on issues where the generated BRTs are fail-to-fail tests. This division of contribution supports the design rationale of \toolname. Further evaluation on Go issues suggests that \toolname has the potential to generalize across programming languages.

In summary, this paper makes the following contributions:

\begin{itemize}[leftmargin=*]
\item We present a preliminary study showing that directly using advanced BRT generators to guide patch generation is not beneficial, motivating the need to rethink how BRTs should be generated and used for issue resolution agents.

\item We propose \toolname, a novel issue resolution agent that guides patch generation by generating multi-faceted BRTs and converting their executions into runtime diagnoses.

\item We conduct an extensive evaluation on two representative benchmarks across five LLM backends, showing that \toolname consistently outperforms strong baseline agents.

\item We publicly release the implementation of \toolname~\cite{SWEDoctor2026}, together with the data used in this paper, to support reproducibility and future research.
\end{itemize}

\section{Motivation}\label{sec:motivation}
Recent studies have developed increasingly effective BRT generation techniques~\cite{eotterpp,issue2test,assertflip}. These techniques are typically evaluated by whether the generated tests can reproduce the reported bugs, and existing studies~\cite{eotterpp,issue2test,assertflip} further use BRTs for \emph{patch validation}, i.e., to check whether a generated patch passes the BRT before submission. 
However, it remains insufficiently understood whether generated BRTs can help the earlier and more central stage of \emph{patch generation} itself. A natural hypothesis is that, if a generated BRT reliably reproduces the reported bug, it can provide the agent with a concrete behavioral target: the agent can generate a patch by trying to make the failing BRT pass. We therefore conduct a preliminary study to examine whether directly using state-of-the-art BRT generators in this way improves patch generation.

\subsection{Preliminary Study}\label{subsec:motivation-design}
\subsubsection{Agent and BRT Generators} We use mini-SWE-agent~\cite{sweagent} as the base agent for our preliminary study. It is a minimal issue resolution agent that follows a simple workflow: it first attempts to generate a BRT for the reported bug, then edits the source code, verifies the patch by re-running the BRT, optionally tests additional cases, and finally submits the patch.
We choose mini-SWE-agent for two reasons. First, it is a representative open-source issue resolution agent and is commonly used as a research baseline~\cite{liveswe,lee2026gistify,tripathy2026swenergy}. Second, its bug-reproduction step is cleanly separable from the workflow, enabling us to easily study how replacing this step with advanced BRT generators affects downstream patch generation.

We consider three recently proposed BRT generators that can represent the current state of the art.

\begin{itemize}
\item \textbf{e-Otter++}~\cite{eotterpp} generates a reproduction test from the issue report, refines it using execution feedback, and improves candidate diversity through issue paraphrasing and multi-mask variation of the code context.
\item \textbf{Issue2Test}~\cite{issue2test} follows a three-phase pipeline: issue and code repository analysis, lint-checked test generation, and execution-driven refinement.
\item \textbf{AssertFlip}~\cite{assertflip} first generates a test that passes on the buggy code, and then flips its assertions to encode the expected behavior, so that the resulting test fails before the fix and passes after the fix.
\end{itemize}

We compare the original mini-SWE-agent with three variants that replace its built-in BRT generation step with one of the three BRT generators, while keeping all other components unchanged. All agents use GPT-5.4 as the backend LLM.

\subsubsection{Dataset} We evaluate them on 100 bug-fixing instances randomly sampled from SWE-bench Verified~\cite{swebench}, a human-validated benchmark widely used to assess the ability of LLMs and agents to resolve real-world software issues.

\subsubsection{Metrics} We use the number of resolved issues (\emph{\#Resolved}) to measure resolution effectiveness. 

To further analyze how BRT generation affects resolution, we divide issues into two categories: issues for which the generated BRT is a fail-to-pass test (denoted as \emph{F$\to$P} issues) and issues for which the generated BRT is a fail-to-fail test (denoted as \emph{F$\to$F} issues).
A generated BRT is a fail-to-pass test if it fails on the original repository, i.e., before the correct patch is applied, but passes on the ground-truth patched repository, i.e., after the correct patch is applied.
In contrast, a generated BRT is a fail-to-fail test if it fails on both the original and ground-truth patched repositories.
For each issue, all the studied generators aim to produce a BRT that fails on the original repository, and generated BRTs that pass on the original repository are discarded. 

For \emph{F$\to$P} and \emph{F$\to$F} issue categories, we report \emph{\#Resolved} separately to examine whether successful bug reproduction translates into improved issue resolution.

\begin{table*}[t]
\centering
\scriptsize
\caption{Issue resolution results of mini-SWE-agent variants equipped with different BRT generators, reported over all issues, \emph{F$\to$P} issues, and \emph{F$\to$F} issues. For each issue category, \#Issues denotes the number of issues; \#Resolved denotes how many of them are resolved by the corresponding variant; $\Delta_{\text{mini}}$ denotes the difference in \#Resolved between the variant and the original mini-SWE-agent on the same issue category.}
\label{tab:prelimi}
\begin{tabular}{l|rrr|rrr|rrr}
\toprule
\multirow{2}{*}{Configuration}
& \multicolumn{3}{c|}{Overall}
& \multicolumn{3}{c|}{\emph{F$\to$P} Issues}
& \multicolumn{3}{c}{\emph{F$\to$F} Issues} \\
\cmidrule(lr){2-4} \cmidrule(lr){5-7} \cmidrule(lr){8-10}
& \#Issues & \#Resolved & $\Delta_{\text{mini}}$
& \#Issues & \#Resolved & $\Delta_{\text{mini}}$
& \#Issues & \#Resolved & $\Delta_{\text{mini}}$ \\
\midrule
w/ e-Otter++  & 100 & 71 & $-3$ & 50 & 34 & $-1$ & 50 & 37 & $-2$ \\
w/ Issue2Test & 100 & 71 & $-3$ & 42 & 33 & $0$  & 58 & 38 & $-3$ \\
w/ AssertFlip & 100 & 73 & $-1$ & 70 & 56 & $+2$ & 30 & 17 & $-3$ \\
\bottomrule
\end{tabular}
\end{table*}

\begin{figure}[t]
  \centering
  \includegraphics[width=0.38\textwidth]{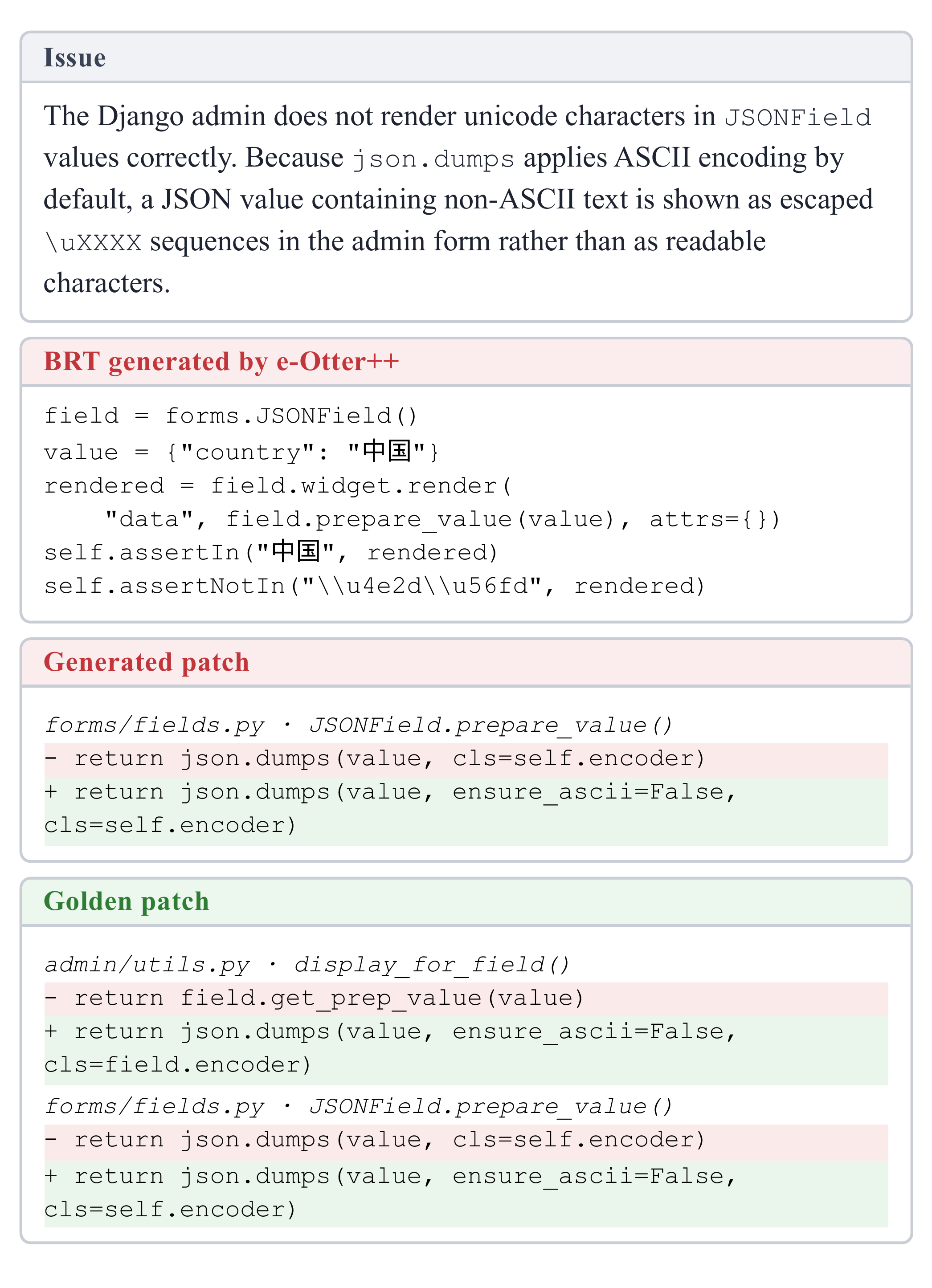}
  \caption{Illustration of the unresolved \texttt{django\_\_django-13512} issue, where a generated BRT covers only the editable form-field path and guides the agent to produce a partial patch.}
  \label{fig:F2Pcase}
\end{figure}

\subsection{Results of Preliminary Study}\label{subsec:motivation-results}
Table~\ref{tab:prelimi} presents the issue resolution results of mini-SWE-agent variants equipped with different BRT generators, reported over all issues, \emph{F$\to$P} issues, and \emph{F$\to$F} issues.

We first compare the overall issue resolution results. The original mini-SWE-agent resolves 74 issues, while the variants equipped with e-Otter++, Issue2Test, and AssertFlip resolve 71, 71, and 73 issues, respectively. This result shows that replacing the built-in reproduction step with advanced BRT generators does not improve end-to-end resolution effectiveness; instead, all three variants resolve fewer issues than the original mini-SWE-agent.

Furthermore, we analyze resolution results separately on \emph{F$\to$P} and \emph{F$\to$F} issues. For each category, we compare each variant with the original mini-SWE-agent on the same issue category. Table~\ref{tab:prelimi} presents the results.

\noindent \textbf{Resolution results on \emph{F$\to$P} issues.}
We first examine \emph{F$\to$P} issues, where the generated BRT fails on the original repository but passes on the patched repository. Intuitively, such BRTs should provide useful guidance for patch generation because they successfully capture a bug-triggering behavior and specify an expected passing behavior after the fix. Surprisingly, however, we find that variants equipped with advanced BRT generators bring only limited improvement. The e-Otter++ variant resolves one fewer issue than the original mini-SWE-agent, and AssertFlip, although producing \emph{F$\to$P} tests for 70\% of the issues, resolves only two additional issues.

To understand why fail-to-pass BRTs help so little, we manually inspect the \emph{F$\to$P} issues that a variant fails to resolve but the original mini-SWE-agent resolves. We find a recurring failure pattern: a reported issue may contain multiple behavioral requirements that a correct patch must satisfy, but the studied BRT generators produce only one BRT for each issue, which may cover only one requirement. When the agent is guided to edit the repository until this single test passes, it can stop at a partial patch that satisfies the tested requirement while leaving other required behaviors unresolved.

For example, \texttt{django-13512} (Figure~\ref{fig:F2Pcase}) reports that a JSON data field containing non-ASCII text, such as Chinese characters, is shown in Django's admin interface with every non-ASCII character escaped into a \texttt{\textbackslash uXXXX} sequence instead of readable text. The root cause is a shared serialization function that escapes non-ASCII characters by default. This function is used by two code paths that render the field: the editable form-field path in \texttt{forms/fields.py} and the read-only admin-display path in \texttt{admin/utils.py}. Therefore, the escaped text appears on both pages, and resolving the issue requires correcting both code paths.

However, the BRT generated by e-Otter++ exercises only the editable form-field path. It renders the form field and asserts that the output no longer contains the escaped sequence. Guided by this single target, the agent edits only \texttt{forms/fields.py}, observes the BRT pass, and submits the patch. The read-only path in \texttt{admin/utils.py} remains unchanged, so the admin display page still escapes non-ASCII text and the issue is not resolved.

This case illustrates a general limitation of directly using fail-to-pass BRTs as patch-generation targets. Passing a single fail-to-pass BRT only shows that the patch satisfies the behavior checked by that test; it does not guarantee that all requirements stated in the issue have been addressed. As a result, the agent may stop at a partial patch that passes the generated BRT but still fails to resolve the full issue, explaining why even valid fail-to-pass BRTs provide limited gains when used as single tests to satisfy.

\noindent \textbf{Resolution results on \emph{F$\to$F} issues.} On \emph{F$\to$F} issues, all variants resolve fewer issues than the original mini-SWE-agent. Specifically, the variants equipped with e-Otter++, Issue2Test, and AssertFlip resolve two, three, and three fewer issues, respectively. This result is expected: on \emph{F$\to$F} issues, the generated tests still fail on the patched repository, indicating that they do not correctly capture the bug-triggering behavior together with its expected fixed behavior. Using such tests as direct patch-generation guidance can therefore mislead the agent and reduce resolution effectiveness.

\noindent \textbf{Insights from our preliminary study.} In summary, our preliminary study shows that directly using generated BRTs as tests to satisfy does not reliably improve patch generation, for two different reasons. The fail-to-pass tests can reproduce the reported bug, but they often capture only partial requirements of the issue. As a result, the agent may generate a partial patch that makes the test pass while leaving other stated requirements unaddressed. This motivates us to generate multi-faceted BRTs that exercise the issue from different requirement-level perspectives, so that patch generation is guided by a more comprehensive view of the reported behavior.

The fail-to-fail tests are not reliable direct targets for patch generation: making such tests pass may lead the agent away from the intended fix. However, this does not mean that their executions are useless. Human developers often face executions that fail in unexpected or imperfect ways, but they do not rely only on the final pass/fail outcome. Instead, they inspect stack traces, runtime states, and executed code paths to localize and understand the fault before fixing it~\cite{zeller,alaboudi}. This suggests that BRT executions should be inspected for runtime evidence rather than used only as pass/fail targets.

\finding{Our preliminary study shows that advanced BRT generators do not automatically translate into better patch generation when each generated test is used directly as a target to satisfy. Fail-to-pass BRTs may reproduce the bug but cover only one manifestation of the reported behavior, leading agents to produce partial patches. Fail-to-fail BRTs are unreliable patch-generation targets and can mislead agents. These results motivate two design directions: generating multi-faceted BRTs for different issue requirements and extracting runtime information from BRT executions rather than relying on pass/fail outcomes alone.}

\section{Methodology}\label{sec:methodology}

\begin{figure*}[t]
\centering
\includegraphics[width=0.93\linewidth]{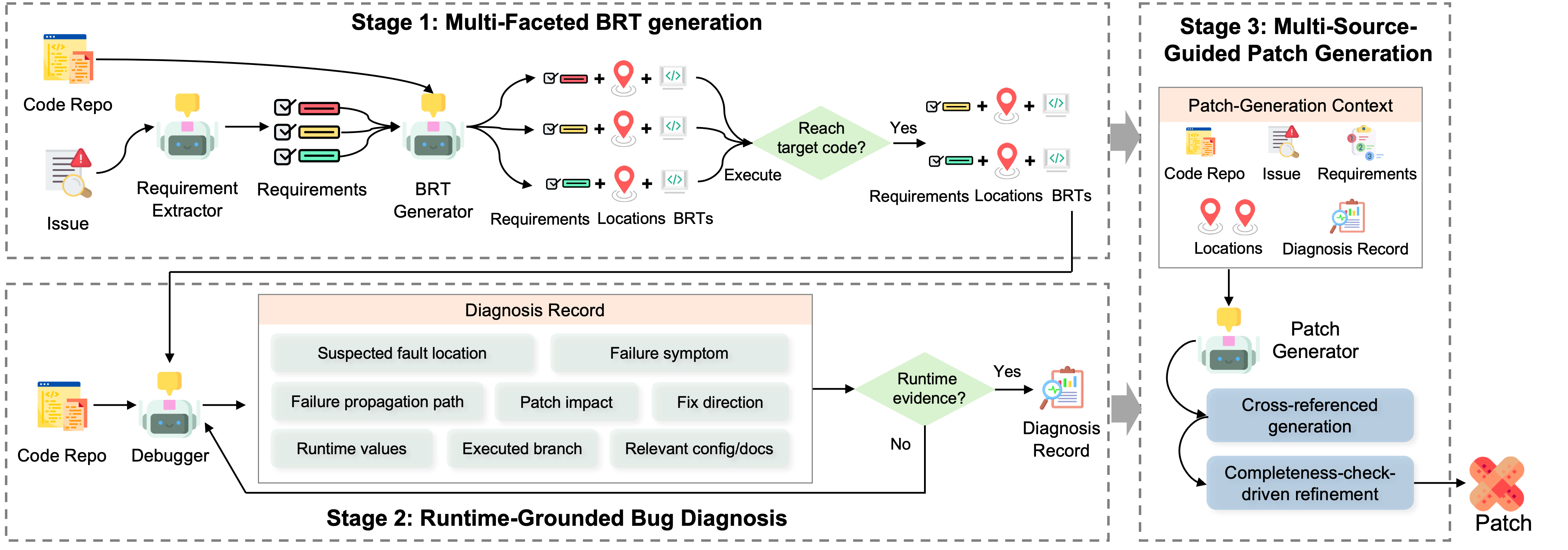}
\caption{Overview of \toolname.}
\label{fig:overview}
\end{figure*}

Motivated by our preliminary study, we propose \toolname, a software issue resolution agent that guides patch generation by generating multi-faceted BRTs and converting their executions into runtime diagnoses.

\subsection{\toolname: In a Nutshell}\label{subsec:overview}
Figure~\ref{fig:overview} gives an overview of \toolname. The input to \toolname is an issue report together with the corresponding buggy repository, and the output is a patch. 

\toolname follows a three-stage pipeline.
First, \textbf{Multi-Faceted BRT Generation} (Section~\ref{subsec:BRT-generation}) extracts the behavioral requirements stated in the issue report and generates targeted BRTs for them. This stage is designed not merely to produce more tests, but to obtain failing executions that expose different facets of the reported behavior. 
Second, \textbf{Runtime-Grounded Bug Diagnosis} (Section~\ref{subsec:rootcause-analysis}) executes the generated BRTs under a debugger and converts the observed runtime behavior into structured diagnosis records. These records summarize resolution-relevant information, such as the suspected fault location, the runtime failure symptom, the failure propagation path, and runtime values observed during debugging.
Third, \textbf{Multi-Source-Guided Patch Generation} (Section~\ref{subsec:patch-generation}) uses the information produced in the previous stages to guide patch generation. It cross-references the diagnosed fault locations with the localization information inferred during BRT generation, and adds a completeness check before submission to reduce partial patches that fix only one observed failure while leaving other stated requirements unaddressed.

\subsection{Stage 1: Multi-Faceted BRT Generation}\label{subsec:BRT-generation}
This stage generates BRTs that later serve as debugging probes. \toolname first extracts the behavioral requirements stated in the issue report and treats them as different facets of the issue. For each facet, \toolname generates a targeted BRT, so that the next stage can diagnose the bug from multiple requirement-specific executions rather than from a single narrow failure. 
The output of this stage is a set of BRTs, each linked to a behavioral requirement and paired with its localized source files and functions.

\noindent\textbf{Multi-faceted requirement extraction.} \toolname first uses an LLM to decompose the issue report into a set of expected-behavior requirements. We define each requirement as an observable behavior stated in the issue report that a correct patch should satisfy, such as an expected output, an exception condition, or a compatibility requirement. Instead of treating this step as free-form issue summarization, \toolname extracts only behaviors grounded in the issue report. This design reduces the risk that the LLM invents plausible but unsupported requirements and helps ensure that the generated BRTs target the behaviors actually requested by the issue.

\noindent\textbf{Targeted BRT generation.}
For each extracted requirement, \toolname uses an LLM to generate a targeted BRT that exercises the behavior described by it. To make test generation focused, \toolname performs requirement-level bug localization before generating the BRT. Given the requirement and the repository, it identifies a ranked list of source files and functions that are likely to be relevant to the requirement. This localization combines deterministic textual evidence, i.e., identifier matches in the repository, with LLM-based localization. The resulting localized context constrains the LLM to reason about code that is more likely to implement the required behavior, making the generated BRT less likely to be a generic restatement of the issue or to exercise unrelated code.

Given the localized context, \toolname generates a BRT through a generate--execute--refine loop. In each iteration, the LLM writes a candidate test, executes it on the buggy repository, and inspects the execution result. The key criterion is not merely whether the test fails, but whether it fails for the behavior specified by the corresponding requirement. When the failure reason is inconsistent with the requirement, \toolname uses the execution feedback and additional code context to revise the test. The loop stops when the generated BRT fails for the intended requirement. This design makes the generated BRTs more targeted and provides the next stage with requirement-specific failing executions to debug.

\noindent\textbf{Execution-based BRT screening.}
A generated BRT is useful for \toolname only if its failure exposes runtime behavior of the code under test. Therefore, after generating a BRT, \toolname screens it by examining whether the failure occurs after the execution reaches the localized code. \toolname keeps a BRT if it fails through an assertion failure or a program exception after exercising the target code, because such executions can provide runtime information for the debugging stage. By contrast, \toolname filters out BRTs that fail before meaningfully reaching the target code, such as failures caused by environment setup errors or missing dependencies, because they provide little evidence about the reported bug.

\subsection{Stage 2: Runtime-Grounded Bug Diagnosis}\label{subsec:rootcause-analysis}
This stage converts the BRTs produced by the previous stage into structured bug diagnoses. A BRT may expose an execution related to a stated requirement, but its pass/fail outcome alone does not explain where the bug lies, which runtime path triggered the observed behavior, or what information should guide the patch. \toolname therefore treats each BRT execution as a debugging entry point: it runs the BRT under a debugger, observes the live program state, and records a diagnosis grounded in runtime observations.

\noindent\textbf{Diagnosis record design.} 
The diagnosis record is designed to make debugging results directly usable for patch generation, rather than passing free-form debugging notes to the patch-generation agent. 

\toolname organizes each diagnosis around various types of patch-relevant information: the suspected fault location, the runtime failure symptom, the failure propagation path, the potential impact of the patch on related code, and a suggested fix direction.
The suspected fault location records the files and functions that are most likely responsible for the bug. The runtime failure symptom records the observed exception, assertion failure, or failure message obtained from the BRT execution. The failure propagation path summarizes how the failure propagates from the failed line back toward the suspected fault location, using the execution stack and code-structure information derived from the abstract syntax tree. The patch impact on related code records whether the suggested change may affect related behavior, such as by changing a suspected function's signature and thereby impacting its callers. The suggested fix direction summarizes, in plain text, how the observed runtime behavior may be addressed. Together, these fields separate where the agent should consider editing, what failure the patch should remove, how the failure arises and propagates during execution, what related behavior the patch should preserve, and how the patch may be constructed.

The record further includes supporting runtime evidence used to ground the diagnosis. This evidence includes debugger expressions and their runtime values, the executed branch when the bug-triggering behavior is observed, and configuration or documentation information related to the suspected code. Together, these fields turn BRT executions into structured, runtime-grounded guidance for patch generation.

\noindent\textbf{Debugger-guided diagnosis.}
To fill the diagnosis record, \toolname uses an LLM debugging agent equipped with a runtime debugger. The design is not tied to a particular debugger: the agent interacts with the repository through an action interface that supports source inspection, shell execution, BRT execution, debugger interaction, and runtime probing. Because Python is a dominant language in widely used SWE-bench-style benchmarks (e.g., the original SWE-bench), we instantiate this interface with PDB~\cite{pdb}, a standard Python debugger. The same interface can be instantiated for other programming languages using their corresponding debuggers.

Given a BRT from the previous stage, the debugging agent runs it on the buggy repository and investigates the resulting execution. The debugger session is kept persistent across commands, because debugging is an interactive process in which later commands depend on earlier observations. The agent uses the debugger to move between stack frames, set breakpoints, re-run the BRT when necessary, and inspect runtime values along the executed path. It then works backward from the observed symptom toward a suspected fault location and produces a conclusion that includes the suspected location, the reasoning behind the diagnosis, and a suggested fix direction. The runtime observations collected during this process are used to fill and ground the diagnosis record.

\noindent\textbf{Runtime-grounding check.}
A key risk in LLM-based diagnosis is that the model may identify a plausible fault location from the issue text and source code without actually inspecting the execution. To prevent such text-only diagnoses, \toolname accepts a diagnosis only when the agent has collected runtime evidence through execution feedback, debugger interaction, or runtime probing. If the agent attempts to conclude before collecting such evidence, \toolname rejects the conclusion and asks the agent to inspect the runtime behavior first. If no runtime evidence is collected during the debugging attempt, the BRT probe yields no diagnosis record. This check ensures that each diagnosis record passed to patch generation is grounded in observed runtime behavior rather than inferred solely from the issue description or source code.

\subsection{Stage 3: Multi-Source-Guided Patch Generation }\label{subsec:patch-generation}
This stage generates the patch using mini-SWE-agent. We choose mini-SWE-agent because it is a minimal issue-resolution agent with a modular and extensible workflow, and it has also been used as the basis for more advanced issue-resolution agents~\cite{liveswe,klearagentforge,swereplay}. \toolname keeps the original agent implementation unchanged and only changes the information provided to the agent before patch generation. Instead of asking the agent to treat a generated BRT as a pass/fail target to satisfy, \toolname constructs a multi-source patch-generation context by cross-referencing the information produced in the previous stages, including the extracted requirements, requirement-level localization, and runtime-grounded diagnosis records, and uses the context to guide both patch generation and pre-submission refinement.

\noindent\textbf{Cross-referenced patch generation.}
\toolname constructs the patch-generation context by combining two complementary sources. The first source is the set of extracted requirements together with their requirement-level localization from Stage~1. This source describes what behaviors the patch should satisfy and where those behaviors are likely implemented. The second source is the diagnosis records produced by Stage~2. This component describes how the observed behaviors arise at runtime, including runtime-grounded evidence such as suspected fault locations, failure symptoms, propagation paths, and relevant runtime values.

Before invoking mini-SWE-agent, \toolname presents these sources as a unified patch-generation context and asks the agent to cross-reference them. The context highlights the behavioral requirements to cover, the suspected edit locations, and the runtime evidence supporting those locations. When the requirement-level localization and diagnosis records point to the same location, \toolname presents that location as a strong candidate for editing. When they diverge, \toolname presents both pieces of evidence and asks the agent to cross-reference them against the issue description, the diagnosis records, and the code before deciding where to edit. This design uses runtime diagnosis to guide patch generation without forcing the agent to follow the suggested locations blindly; the agent can still inspect the repository, compare the evidence, and decide the final edit locations. If no diagnosis record is produced, \toolname falls back to the default mini-SWE-agent workflow.

\noindent\textbf{Completeness-check-driven patch refinement.} 
Even with diagnosis-guided patch generation, the agent may still produce a partial patch that fixes one observed failure while leaving other stated requirements unaddressed. To reduce this risk, \toolname adds a prompt-level completeness check before submission. The agent is asked to revisit the extracted requirements and diagnosis records, verify that each stated behavior has been considered, and reflect on whether the patch is consistent with the runtime diagnosis. This check encourages the agent to further refine the patch when it appears incomplete with respect to either the issue-level requirements or the runtime-grounded diagnosis. Once the patch passes this check, \toolname submits the refined patch.

\section{Evaluation Setup}\label{sec:setup}
\subsection{Research Questions (RQs)}\label{subsec:rqs}
\noindent \textbf{RQ1 (Overall Effectiveness):} How effective is \toolname in resolving software issues compared with existing agents?

\noindent \textbf{RQ2 (Effectiveness across BRT Outcomes):} Can \toolname improve resolution on both \emph{F$\to$P} and \emph{F$\to$F} issues?

\noindent \textbf{RQ3 (Ablation Study):} 
How much do multi-faceted BRT generation and runtime-grounded bug diagnosis each contribute to \toolname's effectiveness?

\noindent \textbf{RQ4 (Stability):} How stable is \toolname's effectiveness across repeated runs compared with existing SE agents?

\noindent \textbf{RQ5 (Cross-Language Generalization):} Can \toolname generalize across programming languages?

\subsection{Benchmark Datasets}\label{subsec:benchmarks}
We evaluate \toolname on two widely adopted benchmark datasets: SWE-bench Verified~\cite{swebench} and SWE-bench Pro~\cite{swebenchpro}. SWE-bench Verified is a popular benchmark for evaluating SE agents, containing 500 real-world human-validated software issues. Since it includes both bug-fixing and feature-request issues, and our study focuses on bug fixing, we follow prior work~\cite{swepolybench} to use only its 436 bug-fixing issues. SWE-bench Pro is a recently proposed benchmark consisting of real-world, complex, and enterprise-level software engineering problems. Compared with SWE-bench Verified, it is more challenging and can better distinguish advanced agents when effectiveness differences on SWE-bench Verified become less pronounced. Similarly, we focus on the 107 bug-fixing instances from Python repositories in SWE-bench Pro. 

\subsection{Baselines}\label{subsec:baselines}
We compare \toolname against two representative open-source software issue resolution agents: mini-SWE-agent~\cite{sweagent} and live-SWE-agent~\cite{liveswe}.

\textbf{mini-SWE-agent} is a lightweight agent that resolves issues by iteratively exploring the repository, generating patches, and validating them with executable feedback. Despite its compact design, it remains a strong open-source agent and is widely used as a research baseline~\cite{liveswe,lee2026gistify,tripathy2026swenergy}.

\textbf{live-SWE-agent} is a self-evolving agent that synthesizes tools during the issue resolution process. It shows leading performance on SWE-bench Verified and SWE-bench Pro~\cite{liveswe}, making it a representative advanced agent for evaluation.

To enable a comprehensive evaluation, we run each agent with multiple LLM backends. We consider five LLMs from four vendors: GPT-5.4~\cite{gpt54}, GPT-5.4-mini~\cite{gpt54mini}, Claude Sonnet 4.6~\cite{claude46}, DeepSeek-V4-Pro~\cite{deepseekv4}, and MiMo-V2.5-Pro~\cite{mimov25pro}.

\subsection{Metric}\label{subsec:metrics}
We use the issue resolution rate (\emph{\%Resolved}) as the primary evaluation metric, defined as the proportion of benchmark issues successfully resolved by an agent.

\subsection{Implementation Details}\label{subsec:implementation}
For Claude Sonnet 4.6, DeepSeek-V4-Pro, and MiMo-V2.5-Pro, we set the temperature to 0 to reduce randomness, which is a common setting in current issue resolution agents~\cite{sweagent, liveswe}. For GPT-5.4 and GPT-5.4-mini, we follow prior work~\cite{liveswe} to use the default temperature setting, because their APIs do not expose temperature control for these models. For all models, we set the reasoning effort to high, because software issue resolution requires multi-step reasoning over issue descriptions, code context, execution feedback, and patch generation.
\section{Results}\label{sec:results}

\subsection{RQ1: Overall Effectiveness}\label{subsec:rq1}
Table~\ref{tab:rq1main} reports the resolution rates of \toolname and the baselines across different LLM backends on SWE-bench Verified and SWE-bench Pro. \toolname achieves the highest resolution rate across all 10 LLM--dataset combinations, i.e., five LLM backends evaluated on two datasets, showing that its improvement is consistent across both model choices and benchmark settings. On average, \toolname resolves 75.7\% of issues on SWE-bench Verified, outperforming mini-SWE-agent and live-SWE-agent by 2.0 and 3.9 percentage points, respectively. The improvement becomes more pronounced on the more challenging SWE-bench Pro benchmark: \toolname resolves 59.4\% of issues, compared with 50.5\% for mini-SWE-agent and 51.4\% for live-SWE-agent, corresponding to gains of 8.9 and 8.0 percentage points, respectively. This larger gain suggests that \toolname is particularly beneficial on more challenging issue resolution tasks.

\begin{table}[t]
\centering
\caption{(RQ1) Issue resolution rates of different agents across LLM backends on SWE-bench Verified and SWE-bench Pro. For each LLM--dataset combination, the highest resolution rate is highlighted in bold.}
\label{tab:rq1main}
\setlength{\tabcolsep}{3pt}
\scriptsize
\begin{tabular}{l| rrr|rrr}
\toprule
\multirow{2}{*}{LLM} & \multicolumn{3}{c|}{SWE-bench Verified} & \multicolumn{3}{c}{SWE-bench Pro} \\
\cmidrule(lr){2-4}\cmidrule(lr){5-7}
& \makecell[r]{mini-SWE\\-agent} & \makecell[r]{live-SWE\\-agent} & \makecell[r]{SWE\\-Doctor} & \makecell[r]{mini-SWE\\-agent} & \makecell[r]{live-SWE\\-agent} & \makecell[r]{SWE\\-Doctor} \\
\midrule
GPT-5.4              & 76.8 & 72.5 & \textbf{77.5}  & 55.1 & 57.9 & \textbf{65.4}   \\
GPT-5.4-mini         & 67.4 & 67.4 & \textbf{70.0} & 44.9 & 40.2 & \textbf{55.1} \\
Claude Sonnet 4.6    & 73.6 & 74.3 & \textbf{75.5}  & 45.8 & 51.4 & \textbf{59.8}  \\
DeepSeek-V4-Pro      & 75.2 & 72.2 & \textbf{77.3}  & 57.9 & 52.3 & \textbf{58.9}  \\
MiMo-V2.5-Pro        & 75.5 & 72.7 & \textbf{78.2} & 48.6 & 55.1 & \textbf{57.9}  \\
\midrule
Average        & 73.7  & 71.8 & \textbf{75.7} & 50.5 & 51.4 & \textbf{59.4} \\
\bottomrule
\end{tabular}
\end{table}

\begin{figure}[t]
\centering
\includegraphics[width=0.45\textwidth]{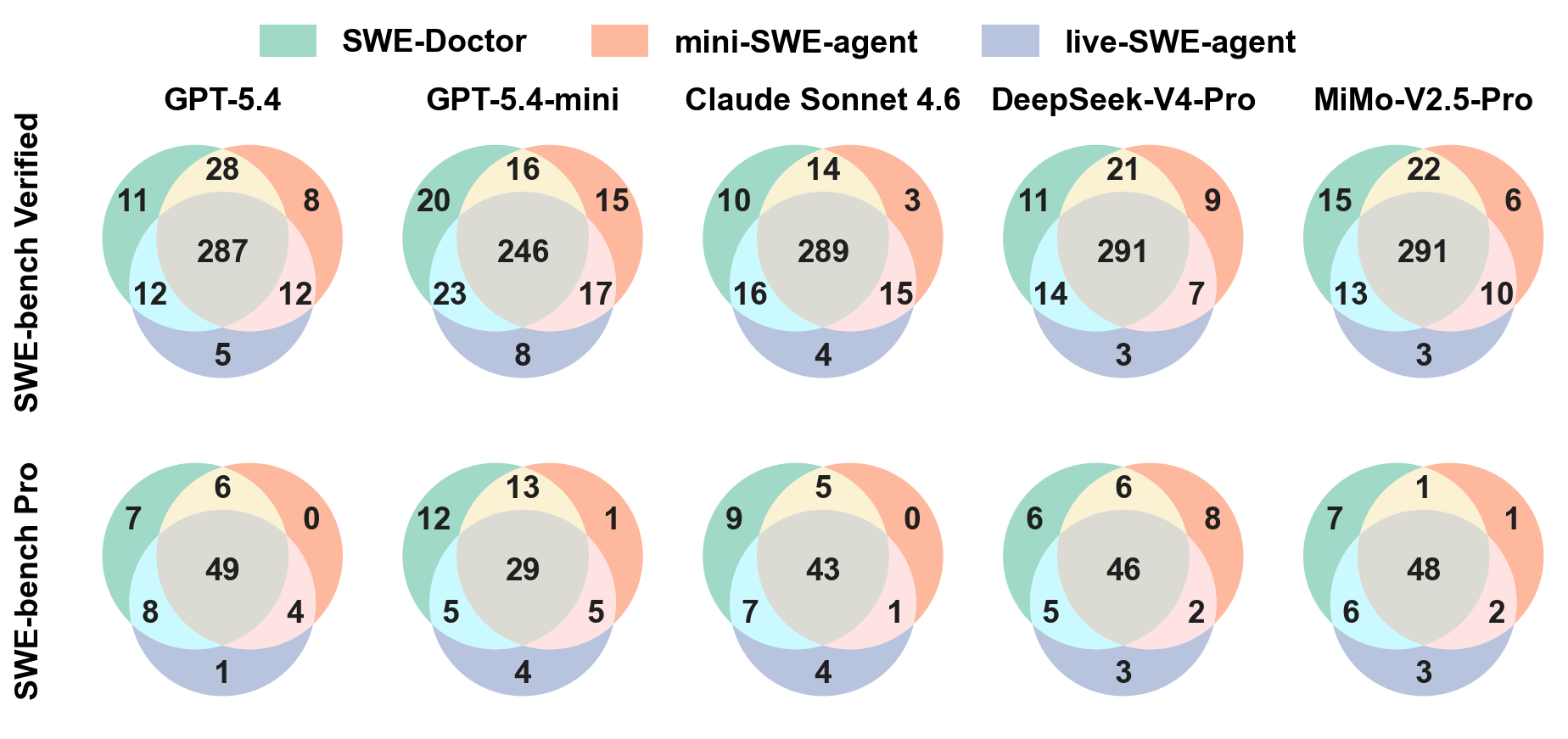}
\caption{(RQ1) Venn diagrams showing the overlap of issues resolved by different agents.}
\label{fig:rq1-venn}
\end{figure}

Furthermore, we analyze the overlap of issues resolved by different agents. Figure~\ref{fig:rq1-venn} shows the Venn diagrams. Overall, \toolname uniquely resolves the largest number of issues, where ``uniquely resolved'' means resolved by one agent but not by the other two. 
Aggregating the counts from individual LLM--dataset combinations, \toolname uniquely resolves 108 issues in total, including 67 on SWE-bench Verified and 41 on SWE-bench Pro. This is more than twice the corresponding aggregated counts for mini-SWE-agent and live-SWE-agent, which uniquely resolve 51 and 38 issues, respectively.
This advantage is consistent: in nine out of the ten LLM--dataset combinations, \toolname uniquely resolves more issues than both baselines. Thus, the improvement of \toolname does not merely come from marginal gains on issues already solvable by existing agents; instead, \toolname expands the set of issues that can be resolved.

We also examine the running cost of different agents. On SWE-bench Verified, \toolname costs \$0.43 per issue on average across LLM backends, compared with \$0.23 for mini-SWE-agent and \$0.22 for live-SWE-agent. On SWE-bench Pro, \toolname costs \$0.64 per issue on average, compared with \$0.52 and \$0.40 for mini-SWE-agent and live-SWE-agent, respectively. Although \toolname incurs higher running cost, the additional cost is accompanied by consistent improvements in issue resolution, especially on the more challenging SWE-bench Pro benchmark.

\finding{\toolname consistently outperforms the baseline agents across all 10 LLM--dataset combinations. The gains are especially pronounced on the more challenging SWE-bench Pro, where \toolname improves the average resolution rate by 8.0--8.9 percentage points over the baselines. Moreover, summed over the LLM--dataset combinations, \toolname uniquely resolves more than twice as many issues as either baseline, suggesting that it expands the set of issues that current agents can resolve.}

\subsection{RQ2: Effectiveness across BRT Outcomes}\label{subsec:rq2}

The preliminary study shows that directly using generated BRTs as target guidance is not beneficial: it can hurt performance on \emph{F$\to$F} issues and provides limited gains even on \emph{F$\to$P} issues. This RQ examines whether \toolname can change this situation. Specifically, we compare the issue resolution rates of \toolname and the original mini-SWE-agent on \emph{F$\to$P} and \emph{F$\to$F} issues across different LLM backends, and also report aggregate results over all backends.

\begin{table}[t]
\centering
\scriptsize
\caption{(RQ2) Issue resolution rates of mini-SWE-agent and \toolname on \emph{F$\to$P} and \emph{F$\to$F} issues across different LLM backends. We also report the aggregate results over all LLM backends. For each issue category, the higher resolution rate is highlighted.}
\label{tab:rq3f2p}
\setlength{\tabcolsep}{3pt}
\begin{tabular}{llrrrr}
\toprule
& & \multicolumn{2}{c}{\emph{F$\to$P} Issues} & \multicolumn{2}{c}{\emph{F$\to$F} Issues} \\
\cmidrule(lr){3-4}\cmidrule(lr){5-6}
Dataset & LLM & \makecell[r]{mini-SWE\\-agent} & \makecell[r]{SWE\\-Doctor} & \makecell[r]{mini-SWE\\-agent} & \makecell[r]{SWE\\-Doctor} \\
\midrule
\multirow{6}{*}{\makecell[l]{SWE-bench\\Verified}}
 & GPT-5.4           & 80.8 & \textbf{83.0} & \textbf{69.6} & 66.9 \\
 & GPT-5.4-mini      & 73.9 & \textbf{77.6} & 60.6 & \textbf{61.5} \\
 & Claude Sonnet 4.6 & 78.6 & \textbf{79.9} & 59.8 & \textbf{63.2} \\
 & DeepSeek-V4-Pro   & 77.9 & \textbf{82.2} & \textbf{72.0} & 71.5 \\
 & MiMo-V2.5-Pro     & 82.6 & \textbf{84.0} & 67.5 & \textbf{71.2} \\
 & \emph{Overall}     & 78.9 & \textbf{81.4} & 66.2 & \textbf{67.2} \\
\midrule
\multirow{6}{*}{\makecell[l]{SWE-bench\\Pro}}
 & GPT-5.4           & 61.0 & \textbf{75.6} & 51.5 & \textbf{59.1} \\
 & GPT-5.4-mini      & 45.5 & \textbf{54.5} & 44.6 & \textbf{55.4} \\
 & Claude Sonnet 4.6 & 56.0 & \textbf{64.0} & 42.7 & \textbf{58.5} \\
 & DeepSeek-V4-Pro   & \textbf{63.6} & 59.1 & 57.1 & \textbf{59.5} \\
 & MiMo-V2.5-Pro     & 66.7 & \textbf{75.0} & 46.3 & \textbf{55.8} \\
 & \emph{Overall}     & 57.1 & \textbf{65.4} & 48.4 & \textbf{57.6} \\
\bottomrule
\end{tabular}
\end{table}

Table \ref{tab:rq3f2p} presents the results. First, \toolname improves issue resolution on \emph{F$\to$P} issues, outperforming the original mini-SWE-agent by 2.5 percentage points on SWE-bench Verified (81.4\% vs. 78.9\%) and by 8.3 percentage points on SWE-bench Pro (65.4\% vs. 57.1\%). Across the 10 LLM--dataset combinations, \toolname achieves higher resolution rates in 9 cases. 
\toolname also improves issue resolution on \emph{F$\to$F} issues, outperforming the original mini-SWE-agent by 1.0 percentage point on SWE-bench Verified (67.2\% vs. 66.2\%) and by 9.2 percentage points on SWE-bench Pro (57.6\% vs. 48.4\%). It achieves higher resolution rates in 8 out of the 10 LLM--dataset combinations. This result contrasts with the motivation study, where directly using fail-to-fail BRTs hurt performance on \emph{F$\to$F} issues. 

\finding{\toolname improves issue resolution on both \emph{F$\to$P} and \emph{F$\to$F} issues. It outperforms mini-SWE-agent in 9 out of 10 LLM--dataset combinations on \emph{F$\to$P} issues and in 8 out of 10 combinations on \emph{F$\to$F} issues. These results show that \toolname can extract useful diagnostic information from BRT executions, even when the generated BRTs fail to reproduce the reported issues.}

\subsection{RQ3: Ablation Study}\label{subsec:rq3}
This RQ evaluates the respective contributions of multi-faceted BRT generation and runtime-grounded bug diagnosis. Due to budget constraints, we randomly sample 50 issues from SWE-bench Pro and use GPT-5.4-mini as the backend LLM. We use this setting because it provides a challenging and discriminative evaluation scenario for ablation analysis: SWE-bench Pro is more challenging than SWE-bench Verified, and Table~\ref{tab:rq1main} shows that, under GPT-5.4-mini, the performance gap between \toolname and the stronger baseline is clearly observable on this benchmark.

We construct four variants of \toolname for the ablation study. To evaluate the effect of multi-faceted BRT generation, we replace this stage with the three advanced BRT generators described in Section~\ref{sec:motivation}, resulting in three variants denoted as \emph{A1--A3}. These variants generate BRTs directly from the issue report without multi-faceted decomposition. To evaluate the effect of runtime-grounded bug diagnosis, we remove this stage and directly use the generated multi-faceted BRTs to guide patch generation, resulting in variant \emph{B}.

\begin{table}[t]
\centering
\caption{(RQ3) Results of the ablation study. We compare the issue resolution rates of \toolname, variants that replace multi-faceted BRT generation with existing BRT generators (A1--A3), and a variant that removes runtime-grounded bug diagnosis (B).}
\label{tab:rq2-ablation}
\setlength{\tabcolsep}{4pt}
\scriptsize
\begin{tabular}{l r}
\toprule
Agent & \%Resolved \\
\midrule
\toolname & \textbf{56.0} \\
\emph{A1}: Stage~1 $\to$ e-Otter++ & 48.0 \\
\emph{A2}: Stage~1 $\to$ AssertFlip              & 46.0 \\
\emph{A3}: Stage~1 $\to$ Issue2Test              & 48.0 \\
\emph{B}: w/o Stage~2 &   48.0 \\
\midrule
mini-SWE-agent                            &  44.0 \\
live-SWE-agent                            &  40.0 \\
\bottomrule
\end{tabular}
\end{table}

Table~\ref{tab:rq2-ablation} presents the results of the ablation study. We first compare \toolname with variants \emph{A1--A3}, where multi-faceted BRT generation is replaced by existing BRT generators. \toolname resolves 56.0\% of the issues, outperforming \emph{A1--A3} by 8.0--10.0 percentage points. 
We then compare \toolname with variant \emph{B}, where runtime-grounded bug diagnosis is removed and the generated BRTs are directly used to guide patch generation. \toolname outperforms \emph{B} by 8.0 percentage points.

Although \emph{A1--A3} and \emph{B} underperform the full \toolname, they still outperform the baseline agents. This suggests that the two stages provide complementary benefits: removing either stage weakens \toolname, but the remaining design still offers useful guidance beyond standard issue-resolution agents.

Recall that our preliminary study identifies two limitations of directly using generated BRTs for patch generation: fail-to-pass BRTs provide limited gains because they cover only one manifestation of the issue, while fail-to-fail BRTs can mislead agents when used directly as patch-generation targets. To further validate whether the two stages of \toolname address these limitations, we compare mini-SWE-agent, variant \emph{B}, and \toolname on \emph{F$\to$P} and \emph{F$\to$F} issues separately.

Table~\ref{tab:rq3-stage} reports the results. On \emph{F$\to$P} issues, variant \emph{B} and \toolname achieve the same resolution rate of 60.0\%, both outperforming mini-SWE-agent. This suggests that the improvement on \emph{F$\to$P} issues mainly comes from multi-faceted BRT generation, supporting our idea that \emph{F$\to$P} issues benefit from covering multiple manifestations of the reported behavior. On \emph{F$\to$F} issues, variant \emph{B} achieves the same resolution rate as mini-SWE-agent, while \toolname performs better than both. This indicates that the improvement on \emph{F$\to$F} issues mainly comes from runtime-grounded diagnosis, which converts failing executions into patch-generation guidance instead of asking the agent to directly satisfy unreliable tests. Together, these results show that the two stages address the corresponding limitations observed in the preliminary study.

\begin{table}[t]
\centering
\caption{(RQ3) Issue resolution rates of mini-SWE-agent, variant \emph{B}, and \toolname on \emph{F$\to$P} and \emph{F$\to$F} issues. The highest rate is highlighted.}
\label{tab:rq3-stage}
\setlength{\tabcolsep}{4pt}
\scriptsize
\begin{tabular}{l rrr}
\toprule
& mini-SWE-agent & \emph{B}: w/o Stage~2 & \toolname \\
\midrule
\emph{F$\to$P} issues & 53.3 & \textbf{60.0} & \textbf{60.0} \\
\emph{F$\to$F} issues & 40.6 & 40.6 & \textbf{53.1} \\
\bottomrule
\end{tabular}
\end{table}

\finding{Both multi-faceted BRT generation and runtime-grounded bug diagnosis contribute substantially to the effectiveness of \toolname. Removing either stage weakens the issue resolution rate by 8.0--10.0 percentage points, while all ablated variants still outperform the baseline agents. Further analysis on \emph{F$\to$P} and \emph{F$\to$F} issues reveals a clear division of contribution: multi-faceted BRT generation mainly contributes to the resolution of \emph{F$\to$P} issues, while runtime-grounded diagnosis mainly contributes to the resolution of \emph{F$\to$F} issues. This alignment supports the design rationale of \toolname.}

\subsection{RQ4: Stability}\label{subsec:rq4}

\begin{table}[t]
\centering
\caption{(RQ4) Issue resolution results over five runs. Avg. and SD denote the mean and standard deviation of resolution rates across runs, respectively. Pass@5 and All@5 denote the proportions of issues resolved in at least one run and in all five runs, respectively. The best value is highlighted for each metric: lowest for SD and highest for the others.}
\label{tab:rq4-randomness}
\setlength{\tabcolsep}{4pt}
\scriptsize
\begin{tabular}{l rrrr}
\toprule
Agent & Avg. & SD & Pass@5 & All@5 \\
\midrule
mini-SWE-agent & 42.4          & 2.2          & 58.0          & 24.0 \\
live-SWE-agent & 45.2          & 3.4          & 62.0          & 24.0 \\
\toolname      & \textbf{54.0} & \textbf{2.0} & \textbf{70.0} & \textbf{40.0} \\
\bottomrule
\end{tabular}
\end{table}

To evaluate the stability of \toolname's effectiveness, we run \toolname and the baseline agents five times under the same setting. Due to budget constraints, as in RQ3, we randomly sample 50 issues from SWE-bench Pro and use GPT-5.4-mini as the backend LLM.

Overall, \toolname maintains its advantage over repeated runs. It achieves an average resolution rate of 54.0\%, higher than the baselines and close to the 55.1\% obtained on the full SWE-bench Pro set with GPT-5.4-mini in RQ1. In terms of variability, \toolname has the lowest standard deviation, at 2.0 percentage points, compared with 2.2 for mini-SWE-agent and 3.4 for live-SWE-agent. This indicates that \toolname's effectiveness remains stable across repeated runs.

Because repeated runs may resolve different issues, we further report two aggregate metrics: Pass@5, the proportion of issues resolved in at least one run, and All@5, the proportion of issues resolved in all five runs. \toolname achieves the highest values for both metrics. Its Pass@5 is 70.0\%, exceeding mini-SWE-agent and live-SWE-agent by 12 and 8 percentage points, respectively. Its All@5 is 40.0\%, exceeding both baselines by 16 percentage points. These results show that \toolname's advantage is consistent across average resolution rate, best-case repeated-run coverage, and consistently resolved issues.

\finding{\toolname maintains stable effectiveness across repeated runs. Over five runs, \toolname achieves the highest average resolution rate among all agents while also showing the lowest variability, with a standard deviation of only 2.0 percentage points. It also achieves the highest Pass@5 and All@5, indicating that \toolname resolves more issues consistently.}

\subsection{RQ5: Cross-Language Generalization}\label{subsec:rq5}
The previous RQs focus on Python issues, because Python is the dominant language in SWE-bench-style benchmarks. For example, SWE-bench Verified contains only Python issues. In this RQ, we examine whether the design of \toolname can generalize beyond Python.

\begin{table}[t]
\centering
\caption{(RQ5) Issue resolution rates of different agents on Go issues from the SWE-bench Pro benchmark. The best result is highlighted in bold.}
\label{tab:rq5-go}
\setlength{\tabcolsep}{4pt}
\scriptsize
\begin{tabular}{l r}
\toprule
Agent & \%Resolved \\
\midrule
\toolname & \textbf{36.0} \\
mini-SWE-agent & 32.0 \\
live-SWE-agent & 30.0 \\
\bottomrule
\end{tabular}
\end{table}

Given our budget constraints, we focus this RQ on Go issues from SWE-bench Pro and use DeepSeek-V4-Pro as the backend LLM. We choose Go because it is the largest non-Python subset in SWE-bench Pro, making it the most representative non-Python language available. We randomly select 50 Go bug-fixing issues from this subset. We choose DeepSeek-V4-Pro because, as shown in Table~\ref{tab:rq1main}, \toolname's advantage over the baseline agents is least pronounced with this backend in the main Python evaluation. This gives a conservative setting for assessing cross-language generalization: if \toolname can still improve issue resolution under this setting, the result provides stronger evidence that its effectiveness is not limited to Python-specific implementation details.

We compare \toolname with mini-SWE-agent and live-SWE-agent under the same setting. To adapt \toolname to Go, we replace the Python debugger PDB with Delve~\cite{delve}, the Go debugger, while keeping the overall design unchanged. This reflects the generality of \toolname's design: when applying it to a different programming language, we only need to replace the language-specific debugger.

Table~\ref{tab:rq5-go} reports the results. \toolname resolves 36.0\% of the Go issues, outperforming mini-SWE-agent and live-SWE-agent, which resolve 32.0\% and 30.0\%, respectively. This result suggests that the advantage of \toolname is not specific to Python: the same design of converting BRT executions into runtime-grounded diagnoses can also improve issue resolution for Go.

\finding{\toolname shows promising cross-language generalization. On Go issues from SWE-bench Pro, \toolname achieves a 36.0\% resolution rate, outperforming mini-SWE-agent and live-SWE-agent, which achieve 32.0\% and 30.0\%, respectively.}

\section{Threats to Validity}
\noindent\textbf{External validity.}
Our evaluation is conducted on two benchmarks, SWE-bench Verified and SWE-bench Pro. Although results on them may not cover all resolution scenarios, both benchmarks are widely adopted and representative for evaluating issue resolution agents. Since Python is a dominant language in SWE-bench-style benchmarks, our main evaluation focuses on Python issues. Nevertheless, the design of \toolname is not tied to Python: the diagnosis stage can be instantiated with language-specific debuggers. To further mitigate this threat, we additionally evaluate on Go issues from SWE-bench Pro, and the results suggest promising cross-language generalization. The choice of baseline agents may also affect external validity. To reduce this threat, we compare \toolname with both mini-SWE-agent, a widely used research baseline, and live-SWE-agent, a recently proposed advanced agent with strong performance on SWE-bench-style tasks. Similarly, the choice of LLM backends may affect the observed results. We mitigate this threat by evaluating \toolname with five representative LLMs from four vendors.

\noindent\textbf{Internal validity.}
LLM-based agents are inherently nondeterministic, so run-to-run variance may affect the comparison between \toolname and the baselines. To mitigate this threat, we conduct repeated runs in RQ4 and quantify the variance across runs. The results show that \toolname maintains its advantage over the baseline agents across repeated runs, suggesting that the observed improvements are not merely due to randomness in a single execution.

\section{Related Work}\label{sec:related-work}
\noindent \textbf{BRT Generation.} LLMs and LLM-based agents have been increasingly used to generate BRTs from issue reports. Early work formulates BRT generation as few-shot or report-to-test synthesis~\cite{libro,plein}, while recent benchmarks such as SWT-Bench~\cite{swtbench} and TDD-Bench Verified~\cite{tddbench} scale the task and provide standardized fail-to-pass and pass-to-pass evaluation protocols, which we also adopt in this paper. Recent approaches further improve reproduction effectiveness through stronger prompting, execution feedback, and agentic control, including Otter~\cite{otter}, e-Otter++~\cite{eotterpp}, Issue2Test~\cite{issue2test}, and AssertFlip~\cite{assertflip}. Industrial evidence also suggests that BRT generation can improve downstream issue resolution in practice~\cite{brtagent}. However, existing studies~\cite{eotterpp,issue2test,assertflip} primarily focus on generating BRTs that successfully reproduce bugs or using BRTs as pass/fail signals for patch validation. In contrast, this paper studies whether generated BRTs can help patch generation itself, and shows that their value depends not only on whether they reproduce a bug, but also on how their executions are used by the issue resolution agent.

\noindent \textbf{Software Engineering Agents.}
Software issue resolution agents, also commonly referred to as software engineering agents~\cite{agentless,liveswe,guo2026eet}, have been increasingly developed with the emergence of benchmarks such as SWE-bench~\cite{swebench}. Existing issue resolution agents typically follow an iterative workflow of repository exploration, code localization, patch generation, and validation. Representative agents include SWE-agent~\cite{sweagent}, mini-SWE-agent~\cite{sweagent}, OpenHands~\cite{openhands}, AutoCodeRover~\cite{autocoderover}, and Agentless~\cite{agentless}, which instantiate this workflow with different degrees of agent autonomy and pipeline structure. Recent work further improves issue resolution through more advanced agent designs, including multi-agent collaboration~\cite{magis,dei}, ensembling and test-time scaling~\cite{traeagent,codemonkeys,sstar}, and self-evolving workflows~\cite{liveswe}. Among these agents, live-SWE-agent~\cite{liveswe} represents a strong recent agent with leading performance on SWE-bench-style benchmarks, so we include it as an advanced baseline in our evaluation. Unlike these agents, which mainly improve how the agent searches, reasons, collaborates, or scales inference, \toolname focuses on how BRTs can be used more effectively during patch generation.
\section{Conclusion}\label{sec:conclusion}
This paper investigates whether and how BRTs can support patch generation in issue resolution agents. Our preliminary study shows that directly using BRTs as pass/fail targets is not beneficial, as fail-to-fail BRTs can mislead agents and fail-to-pass BRTs may still lead to partial patches. Motivated by these findings, we propose \toolname, which generates multi-faceted BRTs, converts their executions into runtime-grounded diagnosis records, and uses the resulting information to guide patch generation. Experiments on SWE-bench Verified and SWE-bench Pro across five LLM backends show that \toolname consistently outperforms strong baseline agents, and further analyses confirm the contributions of both multi-faceted BRT generation and runtime-grounded diagnosis. 

\section{Data Availability} 
The source code of \toolname, together with the data used in this paper, is publicly released \cite{SWEDoctor2026}.

\balance
\bibliographystyle{IEEEtran}
\bibliography{references}

\end{document}